\documentstyle[epsfig]{Przysiezniak-Plenary.1.1}

\begin{document}
\title{Precision Electroweak Measurements}

\author{H.Przysiezniak}
\address{CERN, EP Division\\
1211 Geneva, Switzerland}

\maketitle

\begin{abstract}
This talk describes some of the precision electroweak measurements
from around the world,
namely those related to the Z and W bosons,
the top quark mass, $\sin^2\theta_{\mathrm W}$ at NuTeV,
and three other fundamental measurements: $\alpha^{-1}(m^2_{\mathrm Z})$,
$(g-2)_\mu$ at the E821 BNL experiment as well as the atomic parity violation
(APV) measurement for the Cesium atom. 
These and other measurements are set in the context of the Standard Model
(SM) and of the electroweak fit predictions.
Future prospects for forthcoming experiments are briefly discussed.
\end{abstract}

%

\section*{Introduction}

The motivations to perform precision electroweak measurements today are 
as strong as ever.
Today's generation of experiments now have data of such precision
that the electroweak measurements
are probing the quantum corrections to the SM,
otherwise known as radiative corrections.
These are 
tested by a wide variety of measurements
ranging from the muon magnetic moment
to precision measurements at the Z pole and above in $\mathrm e^+ e^-$
collisions,
as well as precision measurements at hadron colliders.
This talk is not an exhaustive survey
but rather a biased summary of recent and new results,
in particular: those which have an influence on the indirect
determination of the Higgs mass,
and those which are devised to be extremely stringent tests
of the SM.

The measurements that are described here and which enter
the first category are:
the Z line shape and branching ratio measurements 
as well as the Z peak asymmetries 
from which $\sin^2\theta^\ell_{\mathrm eff}$ is extracted,
the W mass,
the top quark mass,
$\sin^2\theta_{\mathrm W}= 1 - m^2_{\mathrm W}/m^2_{\mathrm Z}$ at NuTeV and
$\alpha^{-1} (m^2_{\mathrm Z})$.
The relation between the electroweak quantities are affected by radiative
corrections. The most precisely known quantities being 
$\alpha(m^2_{\mathrm Z})$,
$G_F$ and $m^2_{\mathrm Z}$,
the W mass is related to them as follows
\begin{eqnarray}
m^2_{\mathrm W}= {\pi \alpha(m^2_{\mathrm Z}) \over 
 \sqrt{2} G_F (1-m^2_{\mathrm W}/m^2_{\mathrm Z})
 (1-\Delta r^{\mathrm ew})}
\nonumber
\end{eqnarray}
and the effective weak mixing angle,
$\sin^2\theta^\ell_{\mathrm eff}=(1/4)(1-g^\ell_V/g^\ell_A)$,
by the relation
\begin{eqnarray}
\sin^2\theta^\ell_{\mathrm eff} \cos^2\theta^\ell_{\mathrm eff} =
{\pi \alpha(m^2_{\mathrm Z})\over
\sqrt{2} G_F m^2_{\mathrm Z} (1+\epsilon_1) (1-\epsilon_3/\cos^2\theta_{\mathrm W})},
\end{eqnarray}
where $\epsilon_1$, $\epsilon_3$ and $\Delta r^{\mathrm ew}=f(\epsilon_1,\epsilon_2,\epsilon_3)$
are the radiative corrections.
They are functions of $m^2_{\mathrm top}$ and of 
$\log (m_{\mathrm Higgs}/m_{\mathrm Z})$.
The W mass and effective weak mixing angle measurements
help to constrain the yet unobserved Higgs mass.
Still, it can be deduced
from the above expressions that reducing the error on 
$\sin^2\theta^\ell_{\mathrm eff}$ will constrain
$m_{\mathrm Higgs}$ even more 
particularly if the error on $\alpha (m^2_{\mathrm Z})$
is reduced simultaneously.
The same relationship exists between the W mass and the top quark mass:
reducing
the experimental errors 
simultaneously will help
to constrain the Higgs mass better.

The measurements which enter the second category are:
the Z measurements from which Universality tests are performed, 
fermion pair production and asymmetries from the LEPII $\mathrm e^+ e^-$
collider above the Z pole,
W production and decays,
gauge boson self-interactions and
atomic parity violation.
These measurements stringently test family Universality,
Universality of weak neutral and charged current couplings,
symmetry breaking and radiative corrections. 

\section*{Z Best of Both Worlds}

The measurements described in the following sections
were made at $e^+ e^-$ colliders:
at SLC using the SLD detector,
and at LEP using the ALEPH, DELPHI, L3 and OPAL detectors,
at the Z pole and above in the case of LEPII.
Approximately
$4 \times 10^6$ Zs were accumulated per LEP experiment,
while 555k Zs were taken at SLD using a polarized electron beam.
The measurements
help to constrain the Higgs mass and serve as stringent 
tests of the SM.
The Z electroweak observables (at the Z peak, after QED corrections) are
\begin{itemize}
\item Line Shape:\ \ 
$m_{\mathrm Z},\ \Gamma_{\mathrm Z},\ \sigma_{\mathrm h}^0=12\pi\Gamma_{\mathrm ee}\Gamma_{\mathrm had}/(m_{\mathrm Z}^2 \Gamma_{\mathrm Z}^2)$
\item Branching Ratios:\ \ 
$R_\ell = \Gamma_{\mathrm had}/\Gamma_\ell,\ R_{\mathrm b}=\Gamma_{\mathrm b \bar b}/\Gamma_{\mathrm had}\ {\mathrm and}\ R_{\mathrm c}=\Gamma_{\mathrm c \bar c}/\Gamma_{\mathrm had}$
\item Unpolarized FB Asymmetries for $f=\ell,{\mathrm b},{\mathrm c}$:\ \ 
$A_{FB}^f={\sigma_F^f-\sigma_B^f\over \sigma_F^f+\sigma_B^f}=0.75 A_{\mathrm e} A_{\mathrm f}$
\item Polarization of $\tau$ leptons:
$${\cal P}_{\tau}(\cos\theta)={\sigma_R-\sigma_L\over \sigma_R+\sigma_L}=-{A_{\tau}(1+\cos^2\theta)+2A_{\mathrm e}\cos\theta\over
1+\cos^2\theta+2A_{\tau}A_{\mathrm e}\cos\theta}$$
\item Left-Right Asymmetry:
$$A_{LR}^m={\sigma^f(-|{\cal P}_{\mathrm e}|)-\sigma^f(+|{\cal P}_{\mathrm e}|)\over \sigma^f(-|{\cal P}_{\mathrm e}|)+\sigma^f(+|{\cal P}_{\mathrm e}|)}={\cal P}_{\mathrm e} A^0_{LR} = {\cal P}_{\mathrm e} A_{\mathrm e}\ \ f\neq e$$
\item  Left-Right FB Asymmetries for $f=\ell,{\mathrm b,c,s}$:
$$A_{\mathrm FB}^{\mathrm pol}={\sigma_F^f(-|{\cal P}_{\mathrm e}|)-\sigma_B^f(-|{\cal P}_{\mathrm e}|)-\sigma_F^f(+|{\cal P}_{\mathrm e}|)+\sigma_B^f(+|{\cal P}_{\mathrm e}|)\over \sigma_F^f(-|{\cal P}_{\mathrm e}|)+\sigma_B^f(-|{\cal P}_{\mathrm e}|)+\sigma_F^f(+|{\cal P}_{\mathrm e}|)+\sigma_B^f(+|{\cal P}_{\mathrm e}|)}
=0.75 {\cal P}_{\mathrm e} A_{\mathrm f}$$
\end{itemize}

\subsection*{Z Resonance Parameters at LEP}

The Z resonance parameters
($m_{\mathrm Z}$, $\Gamma_{\mathrm Z}$, $\sigma_{\mathrm h}^0$ and $R_\ell$)
are measured at LEP
using $\mathrm q \bar q$ and $\ell^+\ell^-$ 
event samples collected during scans of the Z peak.
Since the lepton asymmetries $A_{\mathrm FB}^\ell$ are sensitive functions of 
$\sqrt{s}$, they are also extracted from a 
simultaneous fit to the $\mathrm q \bar q$ and $\ell^+\ell^-$ lineshape data.
The SM values are used for the 
${\mathrm Z} \gamma$ and $\gamma$ cross sections 
and the radiatively corrected lineshape functions are fit to data.

Combining the results from the four LEP experiments and for all years,
assuming lepton Universality ~\cite{lepewwgsummary}:
$m_{\mathrm Z}=91.1871\pm .0021\ (.0017{\mathrm syst})$ (SM: 91.18692) GeV,
$\Gamma_{\mathrm Z}=2.4944\pm .0024\ (.0013{\mathrm syst})$ (SM: 2.49589) GeV,
$\sigma^0_h=41.544\pm .037\ (.035{\mathrm syst})$ (SM: 41.4804) nb,
$R_\ell=20.768\pm .024\ (.017{\mathrm syst})$ (SM: 20.7394),
$A^{0,\ell}_{\mathrm FB}=.01701\pm .00095\ (.00060{\mathrm syst})$ (SM: .016342).

The dominant sources of systematic uncertainties are
the normalization of the cross sections, 
the knowledge of the center-of-mass energy $\mathrm E_{cm}$
and 
the QED radiative corrections to the line shape function.

\subsection*{Left-Right Asymmetry at SLD}

This is a powerful, almost systematic free measurement
made using a polarized electron beam 
(${\cal P}_{\mathrm e}$=22\% in 1992; 
63\% in 1993; 73 to 78\% from 1994 to 1997). 
The helicity of the polarized electrons is changed pulse to pulse.
A feedback system keeps the left and right-handed electron currents
equal to $10^{-4}$.
The left and right-handed luminosities are equal to very good
approximation.
The asymmetry is defined as
$A^0_{\mathrm LR}=({1/ {\cal P}_{\mathrm e}})[N_{\mathrm Z}({\mathrm L})-N_{\mathrm Z}({\mathrm R})] / [N_{\mathrm Z}({\mathrm L})+N_{\mathrm Z}({\mathrm R})]=A_{\mathrm e}$
where $N_{\mathrm Z}({\mathrm L,R})$ are the number of 
hadronically decaying Zs counted
with left and right-handed electron beams.
$A_{\mathrm e}$ is a function of 
$\sin^2\theta^\ell$:
$A_{\mathrm e} = [2 (1-4\sin^2\theta^\ell_{\mathrm eff})]/
 [1+(1-4\sin^2\theta^\ell_{\mathrm eff})^2]$,
such that the $A_{\mathrm LR}$ measurement is translated into
an effective weak mixing angle measurement~\cite{lepewwgsummary}
$$A_{\mathrm LR}= .15108\pm .00218
\rightarrow \sin^2\theta^\ell_{\mathrm eff}=.23101\pm .00028\ (.00018{\mathrm syst})\ 
({\mathrm SM}:\ .23145).$$
 The dominant sources of systematic uncertainties
 are the electron polarization and the center-of-mass energy.
As a cross check,
the polarization of the positron beam was measured,
and it was found to be consistent
with zero: ${\cal P}_{\mathrm e^+}=-.02\pm .07\%$.

\subsection*{Z Summed Up}

The lineshape parameters
are used to extract
the partial decay widths ~\cite{lepewwgsummary}:
 not assuming lepton Universality
 $\Gamma_{\mathrm ee}=83.90\pm .12$ MeV,
 $\Gamma_{\mu\mu}=83.96\pm .18$ MeV,
 $\Gamma_{\tau\tau}=84.05\pm .22$ MeV;
 assuming lepton Universality
 $\Gamma_{\ell\ell}=83.96\pm .09$ MeV,
 $\Gamma_{\mathrm had}=1743.9\pm 2.0$ MeV,
 $\Gamma_{\mathrm invisible}=498.8\pm 1.5$ MeV.
Taking the measured value for 
$\Gamma_{\mathrm invisible}/\Gamma_{\ell\ell}=5.941\pm .016$
and dividing it by the SM expectation for
$\Gamma_{\nu\nu}/\Gamma_{\ell\ell}=1.9912\pm.0012$,  
the number of neutrino families is extracted: 
$N_{\nu}=2.9835\pm .0083$.
Converted into a 95\% C.L. upper limit on 
an additional invisible width
assuming $N_{\nu}=3$ gives
$\Delta\Gamma_{\mathrm invisible}<2.0\ {\mathrm MeV}$.

From the leptonic widths
$\Gamma_{\ell\ell}=(1+{3\alpha/ 4 \pi})
({G_F m^3_{\mathrm Z}/ 24\pi\sqrt{2}})
[1+(g_V^\ell/g_A^\ell)^2](1+\epsilon_1)$
and asymmetries
$A_\ell = {2 g_V^\ell g_A^\ell / (g_V^{\ell\ 2} + g_A^{\ell\ 2})}$,
the vector and axial vector coupling ratios 
are determined ~\cite{lepewwgsummary}: 
${g_A^\mu/g_A^{\mathrm e}}=1.0001\pm .0014$,
${g_V^\mu/g_V^{\mathrm e}}=0.981\pm .082$,
${g_A^\tau/g_A^{\mathrm e}}=1.0019\pm .0015$ and
${g_V^\tau/g_V^{\mathrm e}}=0.964\pm .032$,
consistent with Universality of the leptonic weak neutral couplings.

The effective weak mixing angle $\sin^2\theta^\ell_{\mathrm eff}$
is extracted from the SLD and LEP Z-pole leptonic asymmetries.
The asymmetries entering the world average (LEP+SLD)
effective weak mixing angle determination
are presented in Figure 1 ~\cite{lepewwgsummary}.
\begin{figure*}
  \centerline{\epsfig{file=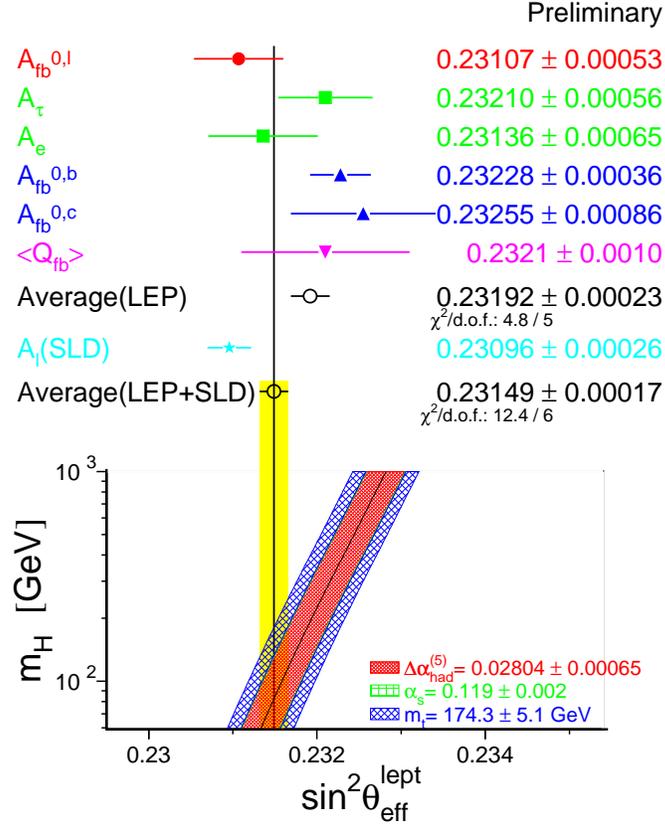,height=11cm}}
\caption[]{Summary of the LEP and SLD asymmetry measurements which enter
in the determination of the world average effective weak mixing angle.}
\end{figure*}
The LEP quark asymmetry measurement
$A_{fb}^{0,{\mathrm b}}$ and
the SLD leptonic asymmetry $A_\ell$ differ by $\sim 3 \sigma$.
This discrepency remains unresolved:
possibly a statistical fluctuation,
or an unkown systematic effect,
or new physics.
Nonetheless.
$\sin^2\theta^\ell_{\mathrm eff}$ 
remains the strongest constraint on the Higgs mass today.

\subsection*{Fermion Pair Production and Asymmetries at LEPII}

Fermion pairs are produced through the radiative return diagram,
when an initial state photon is emitted and the effective center-of-mass
energy $\sqrt{s'}$ is approximately equal to the Z mass,
and through the non-radiative diagram,
when $\sqrt{s'/s}>0.85$. 
The cross section and asymmetries are determined 
for these non-radiative events.
From these measurements, limits on a wide range of physics 
scenarios can be set.

For example,
the cross sections and asymmetries for $\mu^+\mu^-$ and $\tau^+\tau^-$
final states can be used to set limits on contact interactions
between leptons expected to occur in the presence of composite
fermions.
These interactions are
parametrised by an effective Lagrangian which is added to the SM one:
${\cal L}_{\mathrm eff}=[{g^2 \eta / (1+\delta)\Lambda^2}]\sum_{i,j=L,R}\eta_{ij}[\bar {\mathrm e}_i \gamma_\mu {\mathrm e}_i] [{\bar {\mathrm f}}_j \gamma^\mu {\mathrm f}_j]$,
where $g^2 / 4\pi=1$ by convention,
$\eta=\pm$ defines a constructive or destructive interference with the SM,
$\delta = 1(0)$ for $\mathrm f=e(f\ne e)$,
$\eta_{ij}=\pm 1,0$ is the helicity coupling between initial and final state,
$\mathrm e_{L,R}$, $\mathrm f_{L,R}$ are the left and right-handed spinors,
and $\Lambda$ is the scale of the contact interactions.

For all LEP results combined,
the excluded values of $\Lambda$ for models leading 
to large deviations in $\mu^+\mu^-$ and $\tau^+\tau^-$ final states are 
~\cite{lepcontact2000}: 
$\Lambda^{(+,-)}_{\mathrm AA}<(17.6,13.9)$,
$\Lambda^{(+,-)}_{\mathrm VV}<(20.4,17.2)$,
$\Lambda^{(+,-)}_{\mathrm RR}<(12.3,9.7)$ and
$\Lambda^{(+,-)}_{\mathrm LL}<(12.8,10.2)$ TeV,
where $\eta_{LL}=1,1,0,1$, $\eta_{RR}=1,1,1,0$, $\eta_{LR}=-1,1,0,0$
and $\eta_{RL}=-1,1,0,0$
for the AA, VV, RR and LL models respectively.

\section*{W Bosons on Tour}
 
W boson related measurements were performed
by the UA1 and UA2 experiments at the SPS collider
($\mathrm p\bar p$, $E_{cm}=630$ GeV, 
${\cal L}_{\mathrm int}\mathrm \sim 12 pb^{-1}/expt.$),
at the Tevatron by the CDF and D0 experiments
($\mathrm p\bar p$, $E_{cm}=1.8$ TeV, 
${\cal L}_{\mathrm int}\mathrm \sim 120 pb^{-1}/expt.$),
and at LEPII
($\mathrm e^+ e^-$, $E_{cm}=161\rightarrow 209$ GeV, 
${\cal L}_{\mathrm int}\mathrm > 500 pb^{-1}/expt.$).
The W electroweak observables described here are: 
the W production and decays,
the gauge boson self interactions and the W mass and width.
These observables help to constrain the Higgs mass or
serve as stringent tests of the SM.

\subsection*{W Pair Production and Decays at LEPII}

W pairs are produced via the t-channel neutrino exchange
and via the s-channel $\mathrm \gamma-Z$ interference.
The cross section is measured by the four LEP experiments
for all decay channels:
the fully hadronic channel
(4q; BR$\sim$46\%),
the semileptonic channel 
($\mathrm \ell\nu q\bar q$; BR$\sim$43\%),
and the fully leptonic channel ($\ell\nu\ell\nu$; BR$\sim$11\%).
The combined measurements are shown in Figure 2
as a function of the center-of-mass energy ~\cite{lepewwgww}.
The curves correspond to three calculations of the cross section:
the two lower curve calculations (top:YSFWW3; bottom:RacoonWW) 
contain non leading ${\cal O}(\alpha)$ terms, 
whereas the upper curve calculation
(Gentle) does not. The $\chi^2$/dof for data versus Gentle is 11.6/6,
whereas that of data versus Racoon is 5.6/6.
\begin{figure*}[htb!]
  \centerline{\epsfig{file=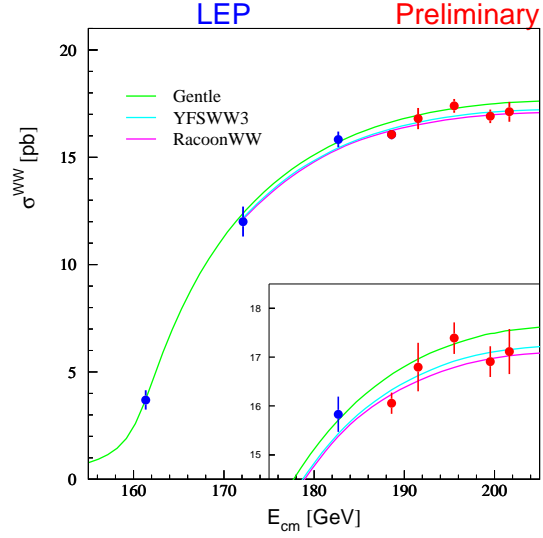,height=8cm}}
\caption[]{LEPII WW cross section as a function of the center-of-mass energy.
The points are the data, the curves are the calculations from 
Gentle (top curve), YSFWW3 (middle) RacoonWW(bottom).}
\end{figure*}

The branching ratio measurements from LEPII
and the Tevatron $\mathrm BR(W \rightarrow e \nu)$ measurement
are consistent with Universality of the leptonic weak charged current
~\cite{lepewwgww,tevatronww}:
 $\mathrm BR(W \rightarrow e \nu) =10.63 \pm 0.20\%$,
 $\mathrm BR(W \rightarrow e \nu) (Tevatron)=10.43 \pm 0.25\%$,
 $\mathrm BR(W \rightarrow \mu \nu) =10.56 \pm 0.19\%$,
 $\mathrm BR(W \rightarrow \tau \nu)=11.02 \pm 0.26\%$.
The result from LEPII assuming lepton Universality is given by:
 $\mathrm BR(W \rightarrow \ell \nu)=10.71 \pm 0.10\%$,
 $\mathrm BR(W \rightarrow q\bar q) =67.85 \pm 0.33\%$.

Using the measured hadronic branching ratio,
the following relation
\begin{eqnarray}
\mathrm {BR_{W\rightarrow qq}\over 1-BR_{W\rightarrow qq}}=(|V_{ud}|^2+|V_{cd}|^2+
|V_{us}|^2+|V_{cs}|^2+|V_{ub}|^2+|V_{cb}|^2)
(1+{\alpha_s(m^2_{\mathrm W})\over \pi})
\nonumber
\end{eqnarray}
and the known values of the other CKM matrix elements,
one finds $\mathrm |V_{cs}|=.993\pm .016$ ~\cite{lepewwgww}. 
This improves by a factor of ten the precision of the PDG value,
which is also obtained without requiring Unitarity of the CKM matrix.

\subsection*{Gauge Boson Self Interactions}

Any deviation from the SM prediction for gauge boson self-interactions
is a true indication of non-SM physics and could indicate W compositeness.
Triply charged (TGC), neutral or quartic gauge couplings 
have been investigated. 
Only the TGCs are discussed here.
The general Lagrangian for TGCs contains 14 parameters.
By requiring C, P and CP invariance and $\mathrm U(1)_{em}$ invariance
($\mathrm q_W=e$), 5 free parameters are left.
Measurements made at LEP at the Z pole set bounds on the couplings.
Finally,
by requiring the additional $\mathrm SU(2)_L\times U(1)_Y$ invariance,
three couplings are left: $\kappa_\gamma$, $g^{\mathrm Z}_1$ 
and $\lambda_\gamma$ where the SM predictions are 1, 1, 0.
$g^{\mathrm Z}_1$  is the coupling strength of the W to the Z.
$\kappa_\gamma$ and $\lambda_\gamma$ define the magnetic moment and the 
electric quadrupole moment of the $\mathrm W^+$:
$\mu_{\mathrm W^+}=({\mathrm e}/ 2 m_{\mathrm W})(1+\kappa_\gamma+\lambda_\gamma)$
and $Q_{\mathrm W^+}=-({\mathrm e}/ m^2_{\mathrm W})(\kappa_\gamma-\lambda_\gamma)$.

\subsubsection*{At Tevatron}

Limits on TGCs are set by looking at 
the $p_T^{\ell\nu}$ distribution of 
$\mathrm WW,WZ \rightarrow (e,\mu) +\nu + jets$ events ($\lambda,\kappa$),
the cross section of 
$\mathrm WZ \rightarrow (e,\mu) +\nu + ee$ events ($\lambda,g_1$),
the $p_T^{\ell}$ distribution of 
$\mathrm WW \rightarrow dileptons$ events ($\lambda,\kappa$),
and 
the $E_T^{\gamma}$ distribution of 
$\mathrm W\gamma \rightarrow (e,\mu) +\nu + \gamma$ events ($\lambda,\kappa$).
Any excess of events is an indication for anomalous TGCs.
Since the cross section with non-SM couplings increases with
$s$, to avoid Unitarity  violation,
the anomalous couplings are expressed as form factors with a scale $\Lambda$
e.g. $\lambda_V(s)=\lambda_V/(1+s/\Lambda^2)^2$.
Under the assumption that the WW$\gamma$ couplings equal the WWZ couplings,
for $\Lambda=2.0$ TeV, 
the one dimensional 95\% C.L. limits are 
$\Delta\kappa_\gamma = (-0.25,0.39)$ and
$\lambda_\gamma = (-0.18,0.19)$ ~\cite{tevtgc},
where $\Delta$ indicates the deviation with respect to the SM prediction.
Assuming the WW$\gamma$ couplings are at the SM value,
for $\Lambda=2.0$ TeV, 
the one dimensional 95\% C.L. limit is 
$\Delta g_1^{\mathrm Z} = (-0.37,0.57)$ ~\cite{tevtgc}.

\subsubsection*{At LEPII}

Limits on TGCs are set 
through measurements of the cross section and 
W production and decay angles
for WW events ($\kappa,g_1,\lambda$),
and through measurements of the cross section, energy and $\theta$ 
of the lepton, jets or $\gamma$,
for single W or single $\gamma$ events ($\kappa,\lambda$). 
The largest sensitivity comes from
the measurement of the W production and decay angles for 
WW events decaying semileptonically.
The one dimensional 95\% C.L. limits for the four LEP experiments combined are 
$\Delta\kappa_\gamma = (-0.09,0.15)$,
$\Delta g_1^{\mathrm Z} = (-0.071,0.024)$ and
$\lambda_\gamma = (-0.066,0.035)$ ~\cite{leptgc}.

\subsection*{W Mass and Width}

\subsubsection*{At Tevatron}

Single Ws are produced in $\mathrm q \bar q$ annihilations of the 
proton and anti-proton valence quarks.
The W hadronic decays are lost in the QCD di-jet background
and thus only leptonic decays are used for the measurement.
The longitudinal information is lost at large $\eta$
due to high background and lack of instrumentation,
such that transverse quantities are used to determine the W mass. 
Two complementary variables are the transverse mass 
$m_T=\sqrt{2 p_T^\ell {\,\,/\hspace{-0.3cm}E_T} 
(1-\cos\phi_{\vec{\,\,/\hspace{-0.3cm}E_T}-\vec{p_T^\ell}})}$
and the transverse momentum of the lepton $p_T^\ell$.
$m_T$ and $p_T^\ell$ are respectively, 
to first order independent and linearly dependent,
of $p^{\mathrm W}_T$.
They are less-more sensitive to the W production process,
and more-less sensitive to detector resolutions.
A Monte Carlo (MC) simulation of the W production, decay and detector
response is used to generate the 
$m_T$ and $p_T^\ell$ distributions as a function
of $m_{\mathrm W}^{true}$.
Today's Tevatron W mass is $m_{\mathrm W}=80.450\pm .063\ (.040{\mathrm stat};.049{\mathrm syst})$ GeV
(CDF: $80.433\pm .079$ ; D0: $80.474\pm .093$) ~\cite{tevatronww}.
The W width is extracted from
a fit to the high end of the $m_T$ distribution by CDF
$\Gamma_{\mathrm W}=2.06\pm .13$ GeV ~\cite{tevatronww}.

The main systematics originate from
the uncertainty on the energy response and resolution,
and on the $p_T$ and ${\,\,/\hspace{-0.3cm}E_T}$ MC modelling.
Since Z events are needed to set the energy-momentum scales and
the W $p_T$ and ${\,\,/\hspace{-0.3cm}E_T}$ models,
the systematic uncertainties are dominated by the limited Z statistics.

\subsubsection*{At LEPII}

The W mass can be extracted via three different methods at LEPII.
The first is a measurement of the W pair production cross section at 
the threshold center-of-mass energy (161 GeV).
At this energy, the cross section is most sensitive to the W mass.

In the second method, the lepton energy spectrum
of the fully leptonic or semileptonic events is measured:
its endpoints are a function of the W mass
$E^\ell_{\pm}=(\sqrt{s}/4)(1\pm \sqrt{1-4m_{\mathrm W}^2/s})$.

Finally, the direct reconstruction is the third and most powerful method.
Only the fully leptonic events are not used here
because they are underconstrained and the invariant mass of the Ws 
cannot be unambiguously determined.
The center-of-mass energy is known very precisely and 
acts as a very strong constraint.
The invariant mass is determined using a five or two constraint
kinematic fit, for 4q or semileptonic events respectively.
In addition, the 4q events require a jet pairing algorithm to determine
which jets correspond to which W. Bad pairings bring about combinatorial
background and contain little or no information on the W mass. 
A convolution or binned likelihood
method is used to extract the W mass from the invariant mass distribution.
Combining the direct reconstruction results from the four LEP experiments gives
$m_{\mathrm W}=80.401\pm .048\ (.027{\mathrm stat};.040{\mathrm syst})$ GeV
(Aleph: $80.449\pm.065$; Delphi: $80.308\pm.090$; L3: $80.353\pm.088$;
Opal: $80.446\pm.065$) ~\cite{lepewwgww}.
The W width and mass are extracted simultaneously in a two dimensional fit
and the LEP combined result for the width is 
$\Gamma_{\mathrm W}=2.19\pm.15$ GeV ~\cite{lepewwgww}.
The main systematics originate from the uncertainties on the LEP 
center-of-mass energy (17 MeV),
the final state interactions (18 MeV), and the fragmentation (28 MeV).

Combining all results from all methods at LEPII with the Tevatron 
and UA2 results, the world average is given by
~\cite{lepewwgplots}
$$m_{\mathrm W}^{\mathrm WA}=80.419\pm .038\ {\mathrm GeV}.$$ 

\section*{Top Mass at Tevatron}

The top quark was discovered at Tevatron in 1995.
Its mass is measured with a precision of 3\%
and makes it the most precisely known quark mass.
The top mass measurement strongly constrains the Higgs mass,
since the radiative corrections are functions of both of these parameters.

Top quark pairs are produced 
through $\mathrm q \bar q$ annihilation into a gluon
which splits into a $\mathrm t \bar t$ pair.
The gluon is not a colour singlet, such that
colour strings are allowed to form
between the top and the anti-top. 
This colour crosstalk is difficult to model and 
is one of the main sources of systematic uncertainties.
The top (anti-top) then decays almost exclusively 
into a b (anti-b) quark and a W boson,
and the W decays either hadronically or leptonically. 

The combined CDF and D0 error on the top mass for 
the all hadronic channel, when both Ws decay hadronically
(BR$\sim$44\%, Signal/Noise$\sim$1/4),
is approximately $8.5\ (7stat)$ GeV.
The invariant mass of the top decay products is determined
in a three constraint kinematic fit.
The di-lepton channel (BR$\sim$5\%, S/N$\sim$4)
has approximately the same precision and
statistical weight.
The events are underconstrained such that no invariant mass can be 
calculated. In this case, the decay dynamics of the data
is compared with the MC expectation 
e.g. using the $p_\nu$ distribution.
The top mass determined from the lepton+jet channel (BR$\sim$30\%, S/N$\sim$1)
has an error of $\sim 5.5\ (4stat)$ GeV,
corresponding to a statistical weight of $\sim$80\% in the average.
A two constraint kinematic fit is used to determine the invariant mass
of the top quark decay products.
The Tevatron top mass, for all channels and for CDF and D0 combined, is
~\cite{tevatronww}
$$m_{\mathrm top}=174.3\pm 5.1\ (3.2{\mathrm stat};4.0{\mathrm syst})\ {\mathrm GeV}$$
with CDF: $176.0\pm 6.5$ GeV and D0: $172.1\pm 7.1$ GeV.
The main systematics originate from the uncertainties on the jet energy scale,
and on the MC modelling of the QCD effects.

\section*{$\sin^2 \theta_{\mathrm W}$ at NuTeV}

The weak mixing angle has been measured by CCFR
and more recently by NuTeV, at Fermilab.
Early determinations of $\sin^2\theta_{\mathrm W}$
gave the SM's successful prediction of $m_{\mathrm W}$ and  
$m_{\mathrm Z}$.
More precise measurements gave the first useful limits on the top quark mass.
The most recent results are the most precise to date and help to constrain 
the Higgs mass.

The Lewellyn Smith relation 
$\mathrm R^\nu=NC/CC=f(\sin^2\theta_W,\chi)$
can be used to extract $\sin^2\theta_{\mathrm W}$.
NC (CC) is the number of neutral (charged) current events,
and $\chi$ is the effect of the sea quark scattering 
e.g. the charm quark mass. 
Unfortunately, $\chi$ is the largest source of systematic uncertainties.
NuTeV rather uses 
the Paschos Wolfenstein relation 
$\mathrm R^- = (NC^\nu - NC^{\bar \nu})/ (CC^\nu - CC^{\bar \nu})
 = (R^\nu - r R^{\bar\nu})/ (1-r)= {1/ 2}-\sin^2\theta_W$
where $\mathrm r=CC^{\bar\nu}/CC^\nu$. 
Using this relation, almost all sensitivity to $\chi$ cancels out 
but separate $\nu$ and $\bar \nu$ beams are needed.
These are supplied by the FNAL 
sign selected quadrupole train which
selects mesons of the appropriate sign.
Contamination of the beams remains small: $<$1/1000 ($<$1/500)
of $\bar\nu_\mu$ ($\nu_\mu$) in the 
$\nu_\mu$ ($\bar\nu_\mu$) beam; 1.3\% (1.1\%) of $\nu_{\mathrm e}$
in the $\nu_\mu$ ($\bar\nu_\mu$) beam.

The NC and CC events are characterized by their event length:
CCs produce long event length muons, whereas
NCs produce short hadronic showers.
For $\nu$ and $\bar\nu$ separately, NuTeV measures
$\mathrm R_{\nu (\bar\nu)}^{meas}=$
$\mathrm [\#\ short\ evts\ in\ \nu_\mu(\bar\nu_\mu)\ mode] /$
$\mathrm [\#\ long\ evts\ in\ \nu_\mu(\bar\nu_\mu)\ mode]$.
$\mathrm R^\nu_{MC}$
and $\mathrm R^{\bar\nu}_{MC}$
are both functions of $\sin^2\theta_{\mathrm W}$ and $\chi$.
The relation
$\mathrm {{\tilde{R}}^-}=R^\nu-x R^{\bar\nu}$  
is minimized
with respect to $\chi$,
giving $x=.5136$.
$\sin^2\theta_{\mathrm W}$ is then varied 
until $\mathrm \tilde{R}_{\mathrm MC}=\tilde{R}_{\mathrm data}$.
For $m_{\mathrm top}=175$ GeV and $m_{\mathrm Higgs}=150$ GeV
~\cite{nutev}
$$\sin^2\theta_{\mathrm W}=1-m^2_{\mathrm W}/m^2_{\mathrm Z}=
.2253\pm.0021\ (.0018{\mathrm stat};.0010{\mathrm syst})\ ({\mathrm SM}:\ .2227)$$
with residual dependence on $m_{\mathrm top}$ and $m_{\mathrm Higgs}$:
$\delta\sin^2\theta_{\mathrm W}=-.00435[(m_{\mathrm t}/175)^2-1]+.00048 \log(m_{\mathrm H}/150)$
with $m_{\mathrm t}$ and $\mathrm m_{\mathrm H}$ in GeV.
This translates into a value of the W mass:
$m_{\mathrm W}=80.26\pm .11\ {\mathrm GeV}.$

\section*{Extra Beautiful Measurements}

A precise determination of $\alpha(m^2_{\mathrm Z})$ becomes 
extremely important in the context of the electroweak fits.
From Equation 1,
one can deduce that the size of $\Delta \alpha(m^2_{\mathrm Z})$ limits
the precision on $\log m_{\mathrm H}$ via the radiative corrections 
$\epsilon_1$ and $\epsilon_3$.

A precise measurement of the muon magnetic moment is also of great
importance.
The relation between the $\mu$ magnetic moment and the spin is given by
$\vec\mu = g ({\mathrm e}/ m) \vec s$.
Radiative corrections bring
deviations from $g=2$, defined as $a_\mu=(g-2)/2$,
but these could also originate from non-SM effects.

Thirdly, the Cesium atomic parity violation measurement is discussed.

\subsection*{In the case of $\alpha^{-1}(m^2_{\mathrm Z})$}

The electromagnetic constant at center-of-mass energy $\sqrt{s}$ 
can be written as ~\cite{davier}
$\alpha(s) = \alpha(0)/ [1-\Delta\alpha_{\mathrm lept}(s)-\Delta\alpha_{\mathrm had}(s)]$.
The leptonic contribution is precisely calculated 
to three loop order: $\Delta\alpha_{\mathrm lept}(m^2_{\mathrm Z})=314.97686 \times 10^{-4}$.
The hadronic vacuum polarization term has the largest uncertainty
and is determined via a dispersion
integral: 
$\Delta\alpha_{\mathrm had}(m^2_{\mathrm Z})=
-[\alpha(0)m^2_{\mathrm Z}/ 3\pi] \cdot Re \int_{4m^2_\pi}^\infty ds 
\{ {R(s)}/ [s(s-m^2_{\mathrm Z})-i\epsilon] \}$, 
where 
$R(s)={\sigma_{\mathrm ee\rightarrow had}/
\sigma_{\mathrm ee\rightarrow \mu^+\mu^-}}$.
$R(s)$ is measured from $\mathrm e^+e^-\rightarrow hadron$ data for
$\sqrt{s}< 40$ GeV,
and is evaluated using perturbative QCD (PQCD) for $\sqrt{s}> 40$ GeV,
giving $\Delta\alpha_{\mathrm had}=(280\pm 7)\times 10^{-4}$.
Summing these results, one obtains
$\alpha^{-1}(m^2_{\mathrm Z})=128.902\pm.090$, 
which in today's electroweak fits translates into
$\Delta \log m_{\mathrm H} = \pm .30$.

However, it is possible to reduce the error on
$\Delta \alpha_{\mathrm had}$
by taking more data at $\sqrt{s} = 0.3-5.0$ GeV,
by using PQCD down to $\sqrt{s}=1.8$ GeV
and by applying theory constraints from QCD sum rules
~\cite{davier2}.
These steps should bring the error down to
$\Delta \alpha^{-1}=\pm .02$,
which would translate into $\Delta \log m_{\mathrm H} = \pm .20$.

$\mathrm e^+ e^-\rightarrow hadron$ 
data are presently being taken by various experiments,
in particular by BESII in China ($2<\sqrt{s}<5\ {\mathrm GeV}$),
and by CMD-2 and SND in Novosibirsk ($0.6<\sqrt{s}<1.4\ {\mathrm GeV}$).
The first run of BESII in 1998 has brought the error on
$R(s)$ from 15-20\% down to 7\% ~\cite{BESII}. 
Combining the Novosibirsk and BESII data,
a 1999 preliminary evaluation
has given $\Delta\alpha^{-1}(m^2_{\mathrm Z})\simeq .035$.
An update will be presented this summer (2000).

\subsection*{In the case of $(g-2)_\mu$ at E821 BNL}

Both experimentally and theoretically,
$a_\mu=(g-2)/2$\  is known with a precision of $\sim 10^{-9}$.
Theoretically, it is senstitive to large energy scales and 
to very high order radiative corrections.
Its precision is limited by second order loop effects 
from the hadronic vacuum polarization.
Experimentally, it is extremely sensitive to new physics.

The theoretical expression
can be written as 
$a_\mu=a_\mu^{\mathrm qed}+a_\mu^{\mathrm weak}+a_\mu^{\mathrm had}$.
As in the case of $\alpha^{-1}$, 
the hadronic contribution has the largest uncertainty.
Today's theoretical calculation gives 
$a_\mu (SM) = (116\ 591\ 594.7\pm 70)\times 10^{-11}$ 
~\cite{leeroberts-lp99}.
Experimentally, $a_\mu$ is being measured at BNL
by the E821 $g-2$ experiment.
A 3.1 GeV $\pi$ beam from the Alternating Gradient Synchrotron
is used. The E821 goal is to achieve a precision of $\pm 40 \times 10^{-11}$.
The present day world average on $a_\mu$ is 
$(116\ 592\ 050\pm 450)\times 10^{-11}$.

The polarized muons from the decay $\pi\rightarrow \mu$ 
move in a uniform $\vec B$ field which is 
$\perp$ to the muon spin ${\vec s}_\mu$ and to the orbit plane.
A quadrupole electric field $\vec E$
is used for vertical focusing.
The spin precession $\omega_s$ minus the cyclotron frequency
$\omega_c$ is given by
${\vec\omega}_a=-({\mathrm e}/ m) \{ a_\mu \vec B 
-[a_\mu - 1/ (\gamma^2-1)]\vec\beta\times\vec E \}$.
The second term in the brace cancels out for
muons with the {\it magic} 
$\gamma_\mu=29.3\rightarrow p_\mu=3.01\ {\mathrm GeV}$.
This is exactly the energy chosen for the experiment.
The decay time spectrum of the positrons from the decay 
$\mu^+\rightarrow e^+ \nu_e \bar\nu_\mu$ is given by
$N_0 e^{-t/\gamma\tau}[1+A(E)cos(\omega_a t +\phi(E))]$.
One counts the number of positrons versus time and fits the above relation to
extract $\omega_a$.

Any deviation from the SM prediction could be interpreted as new physics
e.g. muon sub structure, W compositeness, SuperSymmetry. 

\subsection*{Atomic Parity Violation in Cesium}

The atom is not a purely electromagnetic system.
Parity violation can occur inside the atom.
In order to detect this violation,
the atoms in a gas are first given a preferential handedness e.g.right.
Some suitable property is measured.
The handedness is then reversed e.g.left, and the property is measured again.
If the results of the two measurements differ, then parity is violated.
The left-right asymmetry is expressed as 
$A_{LR}\propto Z^3/ m^2_{\mathrm Z}$,
where $Z$ in the numerator is the number of protons in the atom.
The asymmetry is very small due to the large Z mass in the denominator.
Cesium (Cs) 55 has been chosen because it has a reliable atomic structure
calculation, it has one electron in its outer shell,
and the remaining 54 electrons are tightly bound around the nucleus.
Cs is the simplest of the heavy atoms,
and the heaviest of the simplest.

In the Boulder experiment ~\cite{boulder}, 
a Cs beam passes through a region of perpendicular
electric, magnetic and laser fields.
The highly forbidden 6S$\rightarrow$7S transition occurs in the Cs as the
weak Parity Non Conserving (PNC) transition with a probability  
of $10^{-11}$.
An electric field 
provokes a Stark induced transition which has a probability $10^5$
times larger than the PNC transition and which can interfere with it.
In order to have non zero interference between the two, 
the 6S$\rightarrow$7S transition is excited 
with an elliptically polarized laser field.
The handedness of the region is changed by reversing 
each of the field directions.
The parity violation is apparent
as a small modulation in the 6S$\rightarrow$7S excitation rate
synchronous with all of these reversals.
The modulation is related to the weak charge $Q_W$.
The Boulder result for the weak charge of Cs is ~\cite{boulder2}
$$Q_W\ Cs^{133}_{55}=-72.06 \pm .28_{expt} \pm .34_{th}$$

The SM prediction is given by the relation~\cite{dominici}
$Q_W^{\mathrm SM}=-72.72 \pm .13 - 102 \epsilon_3^{rad} + \delta_N Q_W$
where $\delta_N Q_W$ indicates new physics, and
$\epsilon_3^{\mathrm rad}$ are the radiative corrections
which are evaluated to be $(4.9\pm 1)\cdot 10^{-3}$ 
from high energy data results.
Using this SM prediction and the experimental result,
one finds $Q_W^{expt}-Q_W^{SM}=1.18\pm .46$
or $0.28 \leq \delta_N Q_W \leq 2.08$ at 95\% C.L. ,
which corresponds to a 2.6$\sigma$ discrepency with the SM.
This result can be interpreted in the context of contact interations,
as a measurement of
$\mathrm \Lambda_{LL}$ and $\mathrm \Lambda_{RR}$:
$12.1 \leq \Lambda^+_{\mathrm LL,RR} \leq 32.9 \ \mathrm TeV$.
The LEPII fermion pair cross section and asymmetry measurements
exclude the regions of $\mathrm \Lambda^+_{RR}<12.3$ TeV and
$\mathrm \Lambda^+_{LL}<12.8$ TeV.

\section*{How Fit Is The Standard Model}

Z Universality tests, fermion pairs at LEPII, 
W production and decays, triple gauge couplings,
$a_\mu$ and atomic parity violation measurements 
are all consistent with the SM.
All are stringent tests of the SM
and help to set limits on a wide range of physics scenarios.

Amongst other measurements (see Figure 3),
$\sin^2\theta^\ell_{\mathrm eff}$, $m_{\mathrm W}$,
$m_{\mathrm top}$, $\sin^2\theta_{\mathrm W}$ and 
$\alpha^{-1}(m^2_{\mathrm Z})$ 
enter in the overall electroweak fit 
from which the Higgs mass is extracted.
The $\chi^2/dof$ of the fit is 23/15,
$m_{\mathrm Higgs}=67^{+60}_{-33}\ {\mathrm GeV}$
and $m_{\mathrm Higgs}<188$ GeV at 95\% C.L. .
The direct searches at LEPII give $m_{\mathrm Higgs}>107.7\ GeV$
at 95\% C.L. ~\cite{straessner}.

\begin{figure*}[htb!]
  \centerline{\epsfig{file=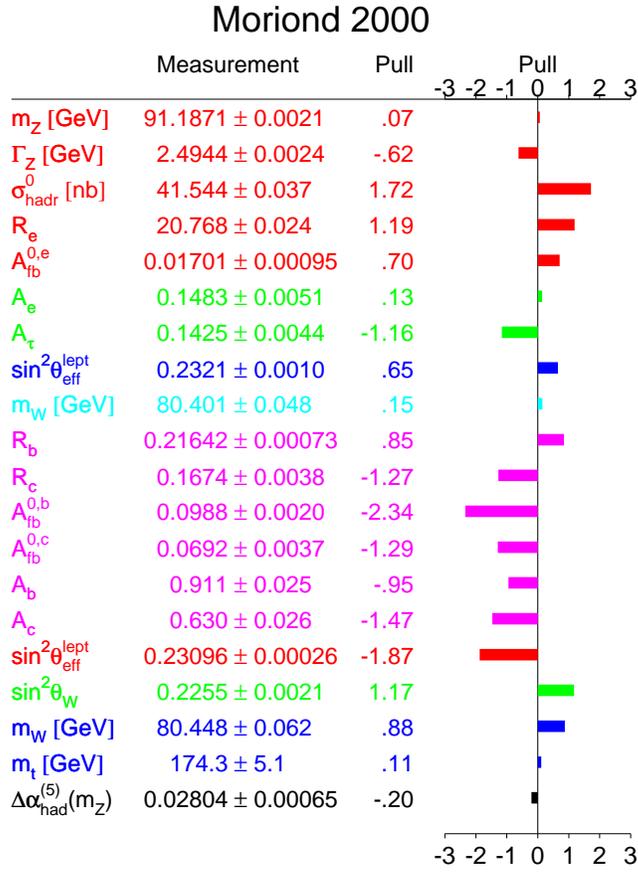,height=13cm}}
\vspace{-1cm}
\caption[]{List of measurements entering the spring 2000 electroweak fit
and pulls with respect to the SM prediction.}
\end{figure*}

The weak mixing angle
($\sin^2\theta^\ell_{\mathrm eff}=.23149\pm.00017$) 
is the strongest constraint
on the Higgs mass today ~\cite{lepewwgsummary}. 
$m_{\mathrm W}$ would be in the race if its error
were $\sim$25 MeV. The W mass measurement confirms the existence of 
the weak radiative corrections to $\sim 7 \sigma$.
Reducing the error on $\alpha^{-1}(m^2_{\mathrm Z})$
will make
the $\sin^2\theta^\ell_{\mathrm eff}$ measurement
an even stronger constraint on the Higgs mass.
The same can be said about the top mass error
with respect to the W mass measurement as a constraint on the Higgs mass.

\section*{Future Prospects}

By the end of LEPII, each experiment will have accumulated 
well over 500 $\mathrm pb^{-1}$. A realistic goal for the W mass measurement
is to attain a world average error of 25 MeV,
and a factor of two improvement on the TGCs.
The goal of the E821 $g-2$ experiment is to reach an error on 
$a_\mu$ of $40 \times 10^{-11}$.
New measurements of $R(s)$ for $\sqrt{s}=0.3$ to 5.0 GeV
at BESII and in Novosibirsk will help to reduce the error
on $\alpha^{-1}$, such that it will most likely reach 
$\Delta\alpha^{-1}\sim .02$.

In the near future (2001), Tevatron will start RUNII.
During the four years of data taking, 
more than 14 $\mathrm fb^{-1}$ per experiment
are expected at $\sqrt{s}=2$ TeV.
The top mass error will come down to 2-3 GeV per experiment,
the overall Tevatron W mass error will be 20-40 MeV.
Approximately 20 $\mathrm fb^{-1}$ will be needed for a $3\sigma$ 
discovery of a SM Higgs of $m_{\mathrm Higgs}<180\ GeV$
~\cite{grannis}.

In the far future, each LHC (ATLAS and CMS) experiment will accumulate 
about 300 $\mathrm fb^{-1}$ over the ten years of running at 
$\sqrt{s}=14$ TeV. The LHC combined experiments
top mass error will be of the order of
2 GeV, the W mass error will be $\leq 25$ MeV per experiment.
Approximately 30 $\mathrm fb^{-1}$ will be needed for a $5\sigma$
discovery of a SM Higgs of $m_{\mathrm Higgs}<1000\ GeV$
~\cite{Fabiola}.

\section*{Conclusion}

The SM is in good shape.
Nothing really anomalous has been observed.
A beautiful future exists for stringent tests e.g. fermion pair production,
TGCs, $a_\mu$ and atomic parity violation.
The same can be said about precision measurements constraining the Higgs
mass e.g. at LEPII, Tevatron and LHC.
There are many good reasons to be optimistic about the future
of precision electroweak measurements !!

\section*{Acknowledgements}

I would like to thank the LEPEWWG, D.Abbaneo, T.Barklow,
A.Blondel, T.Diehl, P.Dornan, A.H\"ocker, K.Grupen, M.Lancaster, B.Pietrzyk,
G.Quast, L.Roberts, D.Schlatter, Z.Zhao
and many more...
An extra special thanks to the CIPANP2000 organizers.

\end{document}